\documentclass{article}
\usepackage[utf8]{inputenc}
\usepackage{amsmath}
\usepackage{amssymb}
\usepackage{amsfonts}
\usepackage{url}

\usepackage{graphicx}

\title{Translation-Invariant Excitons in a Phonon Field}
\author{V.D. Lakhno
\\
Keldysh Institute of Applied Mathematics \\of Russian Academy of Sciences, \\125047 Moscow, Russia \\ lak@impb.ru
}

\date

\begin{document}

\maketitle

\begin {abstract}
Large-radius excitons in polar crystals are considered. It is shown that translation invariant description of excitons interacting with a phonon field leads to a nonzero contribution of phonons into the exciton ground state energy only in the case of weak or intermediate electron-phonon coupling. A conclusion is made that self-trapped excitons cannot exist in the limit of strong coupling. Peculiarities of the absorption and emission spectra of translation invariant excitons in a phonon field are discussed. Conditions when the hydrogen-like exciton model remains valid in the case of electron-phonon interaction are found.
\end{abstract}


\section{Introduction}

The exciton theory is a comprehensive chapter of modern condensed matter physics  \cite{lit1,lit2,lit3,lit4,lit5}. One of its branches is the theory of excitons in polar media  \cite{lit6,lit7}. As in the case of polarons, a description of free excitons in a homogeneous polar medium should be translation invariant (TI). This leads to numerous consequences. Being bosons, excitons, like bipolarons, can form a Bose condensate. However, an experimental confirmation of this possibility has been obtained only recently \cite{lit8}. Some superconductivity theories are also based on the involvement of excitons in the formation of a Bose condensate.

Extensive literature on excitons highlights in detail numerous phenomena concerned with them. For this reason, we will only dwell upon some qualitative differences between the theories of self-trapped excitons in polar media and translation---invariant theory of excitons with electron-phonon interaction (EPI).

 The authors of most modern papers on excitons, when interpreting their spectral lines  (see review  \cite{lit9} and references therein) simply ignore the occurrence of the environment, in particular, the polar medium (in the case of polar crystals). If the influence of the environment is taken into account, a clear picture of exciton spectral lines should be lacking. However, experiments demonstrate clearly distinguishable peaks corresponding to transitions to highly excited states with a very large number of energy level. It seems quite incomprehensible how the presence of an environment and a strong EPI, which should lead to shifting and widening of the exciton transition lines and distortion of the exciton spectrum, leave the nearby lines of transitions to highly excited states distinguishable.

Numerous theoretical investigations of the problem gave rise to the idea of a considerable contribution of EPI into the exciton coupling energy as a result of which a simple hydrogen-like model was replaced by a modified one where the polarization cloud surrounding an electron and a hole, that is, the polaron effect is taken into account by the replacement of the Coulomb interaction by the screened one. The most popular interaction potentials used in interpreting experimental observations are potentials by Haken \cite{lit10}, Bajaj \cite{lit11} Pollmann and B\"{u}ttner \cite{lit12}. In most of the papers, however, the simplest hydrogen-like model is used.

The failure of the model potentials  \cite{lit10,lit11,lit12} can be explained as follows. The point is that the model potentials \cite{lit10,lit11,lit12} were obtained to approximate the exciton coupling energy in the ground state and then they were used to calculate the energy levels in this potential. In fact, it would be appropriate to put each excited state in correspondence with its (self-consistent) potential, for example, the way as it was carried out in \cite{lit13} for F-centers. This problem, however, was not solved in view of its great complexity. Hereinafter, we will show that a solution of this problem is not actually required since in the case of a strong EPI an exact exciton spectrum is hydrogen-like. This explains its widespread use.

Hence, the main result is that in TI systems self-trapped states of excitons caused by EPI are impossible, neither is a self-trapped state of a polaron and a bipolaron  possible \cite{lit14,lit15,lit16,lit17,lit18}. At the same time, as we will show below, the presence of translation invariance leads to some important peculiarities of the TI exciton spectra.

\section{Hamiltonian of an Exciton in a Polar Crystal}

A Hamiltonian of an exciton in a polar crystal is Pekar-Fr\"{o}hlich Hamiltonian, which describes an interaction of an electron and a hole with each other and with optical phonons:
\begin{align}
\hat{H} =&
- \frac{\hbar^2}{2m_1}\Delta_{r_1}- \frac{\hbar^2}{2m_2}\Delta_{r_2}+\sum_k\hbar\omega_{0}(k){\alpha^+_k}{\alpha_k}-\frac{e^2}{\varepsilon_\infty|r_1-r_2|}
\label{eq1}
\\
&+
\sum_k({V_k}e^{ikr_1}{\alpha_k}-{V_k}e^{ikr_2}{\alpha_k} + H.c.),
   \nonumber\\
&V_k=\frac{e}{|k|}\sqrt\frac{2\pi\hbar\omega_0(k)}{V\Tilde{\varepsilon}}, \quad  \omega_0(k)=\omega_0,  \quad \Tilde\varepsilon^{-1}={\varepsilon_\infty^{-1}}- {\varepsilon_0^{-1}}
 \nonumber
\end{align}
where $e$---is an electron charge, $m_1$ and $m_2$---masses of an electron and a hole, $\varepsilon_{\infty}$ and $\varepsilon_0$ are optical and static dielectric permittivities,  $r_1$ and $r_2$---are coordinates of an electron and a hole, $\omega_0 (k)$---is a phonon frequency which in the case of optical phonons is independent of  $k$ and equal to $\omega_0$.

Hamiltonian (\ref{eq1}) corresponds to the case of a continuous polar medium, that is, the case of Wannier-Mott exciton in a polar medium. Different signs in the interaction potential (\ref{eq1}) correspond to different signs of the charge of an electron and a hole.

Having passed in Hamiltonian (\ref{eq1}) from  $r_1$ and $r_2$  to the coordinates of the center of mass  R and relative coordinates  $r$:
\begin{align}
{\mathbf{ r}}_1 = {\mathbf{R}} + \frac{m_2}{M}{\mathbf{ r}},\quad   {\mathbf{ r}}_2 = {\mathbf{R}} - \frac{m_1}{M}{\mathbf{ r}},\quad  M=m_1+m_2,\quad   \mu = \frac{m_1m_2}{M},
    \label{eq2}
\end{align}
we will get:
\begin{align}
\hat{H} =&
- \frac{\hbar^2}{2M}\Delta_{R}- \frac{\hbar^2}{2\mu}\Delta_{r}+\sum_k\hbar\omega_{0}(k){\alpha^+_k}{\alpha_k}-\frac{e^2}{\varepsilon_\infty|\mathbf{ r}|}+
    \label{eq3}
\\
&+
\sum_k{V_k}{\alpha_k}[e^{i\mathbf{k}(\mathbf{R}+m_2\mathbf{r}/M)} - e^{i\mathbf{k}(\mathbf{R}-m_1\mathbf{r}/M)}] + H.c.
\nonumber
\end{align}

Having eliminated in (\ref{eq3}) the coordinates of the exciton center of mass via Heisenberg operator ${S}=\exp(-\frac{i}{\hbar}\sum\hbar {\mathbf{kR}} \alpha^+_k\alpha_k)$ and averaged the Hamiltonian obtained over the wave function of the relative motion $\psi(r)$ we will get:
\begin{align}
{\hat{\bar{H}}} =&
\frac{1}{2M}(\sum_k\mathbf{k}{\alpha^+_k}{\alpha_k})^2+ \sum_k\hbar\omega_0(k)\alpha^+_k\alpha_k + \sum_k[\bar{V_k}\alpha_k + H.c.] + \bar{T} + \bar{U},
\label{eq4}
\\
\nonumber
&\bar{T}
=-\frac{\hbar^2}{2M}\int\psi^*\Delta_r\psi{d^3}{r}, \quad
\bar{U}=-\frac{e^2}{\varepsilon_\infty}\int\psi^*\frac{1}{|\mathbf{r}|}\psi{d^3}r,
\\
&\bar{V_k}
=V_k\langle\psi|\exp(i\mathbf{kr}m_2/M) - \exp(-i\mathbf{kr}m_1/M)|\psi\rangle.
\nonumber
\end{align}

Let us consider different limiting cases for this Hamiltonian.

\section{Exciton Ground State in a Polar Crystal in the Case of Weak and Intermediate Electron-Phonon Interaction}
\label{sec:3}
The contribution of EPI into the exciton energy in the case of weak coupling for  ${m_1}\ne{m_2}$ is nonzero and leads to a reduced energy of an exciton. This automatically follows from expression   (\ref{eq4}), which differs from the case of a polaron by replacements  $V_k\rightarrow\bar{V_k}; m_1,m_2\rightarrow\mu,M$  and addition of constants $\bar{T}$ and $\bar{U}$. As a result, for the ground state energy of a resting exciton in the case of weak and intermediate coupling, according to Lee-Low-Pines  \cite{lit19} we get:
\begin{align}
E=\bar{T}+\bar{U}-\sum_k\frac{|\bar{V}_k|^2}{\hbar\omega_0(k)+\hbar^2k^2/2M}\,.
\end{align}

According to  \cite{lit20}, the value of the ground state energy in the polaron limit of weak coupling when $R_\infty/\omega_0<<1$, where $R_\infty=\mu{e^4}/2\hbar^2\varepsilon_\infty^2$ is equal to:
\begin{align}
E=-&
(\alpha_1+\alpha_2)\hbar\omega_0-R_0\frac{\mu_p}{\mu},
\label{eq6}
\\
\nonumber
&R_0
=\mu{e^4}/2\hbar^2\varepsilon^2_0,\quad \mu_p=\frac{m^p_1m^p_2}{m^p_1+m^p_2},\quad m^p_i=m_i(1+\alpha_i/6),
\\
\nonumber
&\alpha_i=
\frac{1}{2}\frac{e^2u_i}{\hbar\omega_0\tilde\varepsilon}, \quad u_i=\Big(\frac{2m_i\omega_0}{\hbar}\Big)^{1/2},
\end{align}
where $\alpha_i,i=1,2$  are EPI constants for an electron and a hole, respectively.

Formula (\ref{eq6}) has a simple meaning. Let us consider free electron and hole polarons with energies  $-\alpha_1\hbar\omega_0$ and $-\alpha_2\hbar\omega_0$ . At large distances between them, there must be a residual Coulomb interaction. For an electron, it is produced by a polarization charge  $e_{ind}=-e/\tilde\varepsilon$, induced in a polar medium by a hole and a hole's Coulomb field created by its effective charge $e_{eff}=e/\varepsilon_\infty$. As a result, there will be an attractive potential between the electron and the hole, induced by the total charge  $e_{tot}=e_{ind}+e_{eff}=-e/\tilde\varepsilon+e/\varepsilon_\infty=e/\varepsilon_0$. This attractive potential leads to a contribution to the energy, determined by the second term on the right-hand side of  (\ref{eq6}), which is the effective Rydberg, which involves  $\varepsilon_0$ and a relative mass of the electron and hole polarons  $\mu_p$, determined by the polaron masses of the electron and hole.

From (\ref{eq6}) it follows that in the absence of a static Coulomb interaction between an electron and a hole $(\varepsilon_0=\infty$) there is an ordinary polaron shift in the energies of an electron and a hole moving independently. In another limiting case, when EPI is absent $\tilde\varepsilon=\infty$, \mbox{from (\ref{eq6})} follows the ordinary expression for the effective hydrogen atom in the ground state.

In the region of an intermediate coupling constant the phonon contribution in the ground state exciton energy for some crystals was calculated by Gerlach and Luczak \cite{lit20}. For example, for CuCl crystal with $\varepsilon_0=7.4, \varepsilon_\infty=3.7, m_1=0.44 m_0, m_2=3.6 m_0, \alpha_1=2.005, \alpha_2=5.735, \hbar\omega_0=27.2$ meV, where $m_0$ is the mass of free electron in vacuum, it was obtained $E=443.91$ meV for exciton in phonon field and $E=389.6$ meV without phonon input. Thus, the input of phonons to the ground state exciton energy in the region of intermediate coupling can be significant. Note that no any ``phase transition'' between weak and strong coupling limits occur in the region of intermediate coupling (see also Section 
 \ref{sec:7}).

\section{Exciton Ground State in a Polar Crystal in the Case of Strong Electron-Phonon Interaction}

Hamiltonian (\ref{eq4})  is independent of the coordinates of the exciton center of mass  $\mathbf{R}$. Therefore, trapping of an exciton, that is, the formation of an exciton localized in  $\mathbf{R}$-space is impossible.  The reason is that the total momentum of an exciton commutates with the Hamiltonian, accordingly, the eigen wave functions of the Hamiltonian are, at the same time, the eigen wave functions of the total momentum operator $\hat{P}$ (Section \ref{sec:5}), that is, plane waves in $\mathbf{R}$-space.

Another important conclusion resulting from the form of  (\ref{eq4}), is that in the limit of strong coupling with phonons an exciton is not polarizing. In other words, in the limit of strong EPI an exciton behaves like a free exciton in a nonpolar medium.

Let us show this for the case when $\omega_0(k)$ is independent of  $k$. In this case Hamiltonian  (\ref{eq4}) has the same structure as the bipolaron one considered in  \cite{lit14,lit15,lit16,lit17,lit18}. Repeating the calculations made in these papers for a bipolaron in the strong coupling limit we derive from  (\ref{eq4}) the ground state energy in the form:
\begin{align}
E=\Delta{E}+2\sum_k\bar{V_k}f_k+\sum_kf^2_k+\bar{T}+\bar{U},
\label{eq7}
\end{align}
\begin{align}
|\psi(r)|^2=(2/\pi{l^2})^{3/2} \exp(-2r^2/l^2),
\label{eq8}
\end{align}
\begin{align}
f_k=\pm{c}(V_k/\hbar\omega_0) \exp(-k^2/a^2),
\label{eq9}
\end{align}
where the sign ``+'' in (\ref{eq9})  for $f_k$ refers to the case of  $m_1<m_2$, while the sign ``$-$'' to the case of $m_1>m_2$;  $a,l,c$  are variational parameters involved in variation functions  $\psi$ and $f_k$, $\Delta{E}$---is the so-called recoil energy  \cite{lit14,lit15,lit16,lit17,lit18}.

Substituting $(\ref{eq8})$, $(\ref{eq9})$  into  $(\ref{eq7})$ we express the ground state energy as:
\begin{align}
E=&
0.633\frac{\hbar^2a^2}{M}-\frac{e^2}{\sqrt{2\pi}\tilde\varepsilon{a}} \Big( \frac{1}{\sqrt{\frac{l^2m^2_2}{8M^2}+\frac{1}{a^2}}}-\frac{1}{\sqrt{\frac{l^2m^2_1}{8M^2}+\frac{1}{a^2}}} \Big)^2+
\nonumber
\\
\label{eq10}
&+\frac{3\hbar^2}{2\mu{l^2}}-2\sqrt{\frac{2}{\pi}}\frac{e^2}{\varepsilon_{\infty}l}\ ,
\end{align}
where minimization in  $c$  is already performed.

It should be noted that expression  $(\ref{eq10})$ yields a solution of the two-particle problem with different masses in the case of repulsion between the particles also, if we replace the sign ``$-$'' by ``+'' in the parentheses in  $(\ref{eq10})$ and in front of the last term in the right-hand side of  $(\ref{eq10})$. In this case for $m_1=m_2$ and  $a=8/(\sqrt{2}l)$   this expression is transformed into that obtained for a bipolaron in  \cite{lit14}.

Expression $(\ref{eq10})$ is obtained for the case of strong coupling when  $a\rightarrow\infty$. It can be shown, however, that the function  $E=E(l,a)$ does not have a minimum in this limit. The only minimum which  $E(l,a)$ has corresponds to the values: $l=\frac{3\sqrt{\pi/8} \, \hbar^2\varepsilon_\infty}{\mu e^2}, \quad a=0, \quad E_{min}=-(4/3\pi)\mu e^4/(\hbar^2\varepsilon^2_\infty)$, which correspond to the case of a free exciton.

Hence, our initial assumption made in deriving  $(\ref{eq10})$, about the existence of a phonon contribution into the exciton energy in the case of strong EPI appeared to be erroneous. The result obtained suggests that the phonon contribution into the exciton energy which corresponds to a polarizing exciton can be nonzero only for finite values of $\alpha$, that is, for finite values of the EPI constant (Section \ref{sec:3}).

It follows that for rather large values of the EPI constants of electrons and holes when the energy of a polaronic exciton is close to the energy of a free exciton, it may become energetically more advantageous for an exciton to decay into two independent polarons with energies  $E_p^e$ and  $E_p^h$ for an electron and a hole, respectively.

The conditions of the exciton stability with respect to this decay is fulfillment of the inequality:

\begin{align}
|E^{\text{exc}}|\geq|E_p^e|+|E_p^h|.
\label{eq11}
\end{align}

In the case of strong coupling with the use of expressions for the energy of a free exciton  $E^{\text{exc}}=-\mu{e^4}/2\varepsilon^2_\infty\hbar^2$  and that of TI polarons:   $E_p^{e,h}=-0.06286 \times m_{1,2} e^4/\tilde{\varepsilon}^2\hbar^2$  \cite{lit14,lit15,lit16,lit17,lit18} the region of stability  of an exciton, according to  $(\ref{eq11})$ will be determined by the condition:
\begin{align}
0.5-0.5\sqrt{1-0.5\varepsilon^2_\infty/\tilde{\varepsilon}^2} < m_{1,2}/M < 0.5+0.5\sqrt{1-0.5\varepsilon^2_\infty/\tilde{\varepsilon}^2}.
\label{eq12}
\end{align}

It follows from  $(\ref{eq12})$  that, for the condition of the exciton stability to be fulfilled, the value of the static dielectric permittivity  $\varepsilon_0$ should be less than  $3.4{\varepsilon_\infty}$.

Let us pay attention to the fact that functional $(\ref{eq10})$ does not transform into the functional for the F-center when any of the masses tends to infinity, since such a transition would correspond to the loss of translational invariance of the initial system. As is shown in \cite{lit16}, a free TI polaron will be captured by the Coulomb attractive charge of the F-center, only for a certain critical value of the static dielectric constant. In the case of free electron and hole TI polarons, such capture with the formation of an exciton state will occur only if condition $(\ref{eq12})$ is fulfilled.

\section{Spectrum of a TI Exciton}
\label{sec:5}
To find the spectrum of Hamiltonian  $(\ref{eq3})$ we will seek a solution of problem  $(\ref{eq3})$ in the form:
\begin{align}
\Psi=|\psi(r)\rangle|X(R,\{\alpha_k\}).\rangle
\label{eq13}
\end{align}

Then the average value of Hamiltonian $(\ref{eq3})$  with respect to  $|\psi(r)\rangle$  will have the form:
\begin{align}
&{\hat{\bar{H}}}=
\langle\psi|\hat{H}|\psi\rangle=-\frac{\hbar^2}{2M}\Delta_R + \sum_k{\hbar\omega_0(k)}{\alpha^+_k}{\alpha_k}
\nonumber
\\
&+ \sum_k{\bar{V_k}}[{e^{ikR}}{\alpha_k+H.c.}]+\bar{T}+\bar{U},
\label{eq14}
\end{align}
which coincides with the polaron Hamiltonian with an accuracy of constants  $\bar{T}$  and $\bar{U}$ and the replacement of $V_k$ by $\bar{V_k}$ , determined by (\ref{eq4}). Below we will believe  $\hbar=1$.

Following \cite{lit21}, we will choose the wave function $|X\rangle$, involved in  (\ref{eq13}), in the translation invariant form:
\begin{align}
&|X(\mathbf{P})\rangle=\Big[C_P{e^{i\mathbf{PR}}} +
\nonumber
\\
&+\sum^\infty_{N=1}\sum_{k_1\ldots{k_N}}C_{P,k_1\ldots{k_N}} \cdot e^{i(\mathbf{P}-\mathbf{{k}_1}-\mathbf{k_2}-\ldots-\mathbf{k_N})\mathbf{R}}\alpha^+_{k_1}\alpha^+_{k_2}\ldots\alpha^+_{k_N}\Big]|0,\rangle
\label{eq15}
\end{align}
where: $C_P$ and $C_{p,k_1\ldots{k_N}}$---are normalized constants, $|0\rangle$---is a vacuum wave function, $\mathbf{P}$---is a vector of eigen values of the total momentum operator:
\begin{align}
\hat{P}=-i\partial/\partial\mathbf{R} + \sum^\infty_{i=1}\mathbf{k_i}\alpha^+_{k_i}\alpha_{k_i}.
\label{eq16}
\end{align}

Since the total momentum operator  (\ref{eq16}) commutates with Hamiltonian  $\bar{H}$,  the wave function $|X(\mathbf{P})\rangle$ is at the same time their eigen wave function:
\begin{align}
&{\hat{\bar{H}}}|X(\mathbf{P})\rangle=E(\mathbf{P})|X(\mathbf{P})\rangle,
\\
\nonumber
&\hat{P}|X(\mathbf{P})\rangle=\mathbf{P}|X(\mathbf{P}).\rangle
\label{eq17}
\end{align}

Let $|X(\mathbf{P})\rangle$ be the wave function of the ground state. Then, according to  \cite{lit21}, the wave function of the excited one-phonon state $|\psi(\mathbf{K}_j)\rangle$:

\begin{align}
|\psi(\mathbf{K}_j)\rangle=\alpha^+_{k_j}|X(\mathbf{P}),\rangle
\end{align}
where $\mathbf{K}_j$ has the meaning of the total momentum in the j-th excited state, will possess the properties:
\begin{align}
&\hat{P}|\psi(\mathbf{K}_j)\rangle=\mathbf{K}_j|\psi(\mathbf{K}_j)\rangle=(\mathbf{P}+\mathbf{k}_j)|\psi(\mathbf{K}_j)\rangle,
\\
\nonumber
&{\hat{\bar{H}}}|\psi(\mathbf{K}_j)\rangle=\varepsilon(\mathbf{K}_j)|\psi(\mathbf{K}_j)\rangle=(E(\mathbf{P})+\omega_{k_j})|\psi(\mathbf{K}_j)\rangle=
\\
\nonumber
&=(E(\mathbf{K}-\mathbf{K}_j)+\omega_{k_j})|\psi(\mathbf{K}_j).\rangle
\end{align}

Hence, the spectrum of the excited states has the form:
\begin{align}
\varepsilon(\mathbf{K})=E(\mathbf{K}_j-\mathbf{k}_j)+\omega_0(\mathbf{k}_j),\quad\omega_0(\mathbf{k}_j)=\omega_{k_j}.
\end{align}

For the quantity  $E(\mathbf{K}_j-\mathbf{k}_j)$ it was shown in  \cite{lit21} that:
\begin{align}
E(\mathbf{K}_j-\mathbf{k}_j)\leq{E}(0)+\frac{(\mathbf{K}_j-\mathbf{k}_j)^2}{2M}.
\label{eq21}
\end{align}

Actually, according to  \cite{lit22}, in this case instead of inequality  $(\ref{eq21})$  the exact equality takes place and for $\mathbf{K}_j=0$ the spectrum has the form:
\begin{align}
\varepsilon(\mathbf{k}_j)=E(0)+\omega_0(\mathbf{k}_j)+\mathbf{k^2_j}/(2M).
\label{eq22}
\end{align}

It should be noted that in the general case the wave function of an excited state containing N phonons is written as:
\begin{align}
|\psi_{k_1,\ldots,k_N}\rangle=\alpha^+_{k_1}\alpha^+_{k_2}\ldots\alpha^+_{k_N}|X(\mathbf{P}),\rangle
\end{align}
for which the inequality holds:
\begin{align}
\varepsilon({k_1,\ldots,k_N})\leq{E}(0)+\sum^N_{i=1}\omega_0(\mathbf{k}_j)+\frac{(\mathbf{K}-\mathbf{k}_1-\ldots-\mathbf{k}_N)^2}{2M},
\end{align}
where $\mathbf{K}$---is the total momentum corresponding to N phonon excitations.

It should also be noted that in the case of an exciton, when there is a set of electron excitations numbered by index  n (n can have the meaning of a set of quantum numbers),  $(\ref{eq22})$ takes the form:
\begin{align}
&\varepsilon_1(\mathbf{k}=0)=E_1(0)=E^{\text{exc}}
\label{eq25}
\\
\nonumber
&\varepsilon_n(\mathbf{k}\neq{0})=E_n(0)+\omega_0(\mathbf{k})+\mathbf{k}^2/2M,\quad n=1,2,\ldots
\end{align}

\section{Peculiarities of Light Absorption and Emission by TI Excitons}

Let us consider the case of optical phonons when  $\omega_0 (\mathbf{k})$ is independent of  $\mathbf{k}$, that is, the case of polar crystals. For direct excitons, according to  $(\ref{eq25})$, in addition to an ordinary discrete spectrum $E_n (0)$ there is a quasi-continuous spectrum with energies  $E_n (0)+\omega_0+k^2/2M$, which makes spectrum  $E_n (0)$ discernable only when  $\omega_0>|E_1 (0)-E_2 (0)|$.  If the condition:

\begin{align}
|\varepsilon_1-\varepsilon_{n_{c}+1}|>\omega_0>|\varepsilon_1-\varepsilon_{n_c}|
\label{eq26}
\end{align}
is fulfilled, only  $n_c$ first levels of an exciton will be discernible. This result can be used to investigate soft phonon modes concerned with structure phase transitions in crystals, for example, in cuprate superconductors. Thus, if away from a phase transition condition $(\ref{eq26})$ is fulfilled for $n_c=2$, then an optical transition of an exciton from the ground state to the first excited one will make a contribution into the absorption. At the point of the phase transition when  $\omega_0\approx0$ this contribution will be lacking since all discrete levels of an exciton $E_n (0$) fall into the quasi-continuous spectrum.

As shown in \cite{lit14,lit15,lit16,lit17,lit18}, the excitation spectrum $(\ref{eq25})$ can be treated as a spectrum of renormalized EPI phonons, which represent the initial phonon with which an electron and a hole are associated. Scattering of light by such phonons with frequency  $\nu$ will give rise to satellite frequencies $\nu^\text{e}_{n,k,+}=\nu+|\varepsilon_n(k)|$  and $\nu^\text{e}_{n,k,-}=\nu-|\varepsilon_n(k)|$ in the scattered light. Hence, depending on the values of the parameters involved in these expressions, the absorption/emission spectrum of a TI exciton can be more complicated than that of a free exciton. For example, when condition  $(\ref{eq26})$ is fulfilled with  $n_c=2$  absorption (emission) of light can take place without changes in the main quantum exciton number n. In this case the absorption (emission) curve will have a characteristic double-peaked distribution of intensity with maxima for $\nu^\text{exc}_{1,\pm}\approx\nu\pm\omega_0$ \cite{lit23}.

{
Polaron character of excitons was also observed in such hybrid organic-inorganic semiconductors as two-dimensional lead-halide perovskites by implementing high-resolution resonant impulsive stimulated Raman spectroscopy (RISRS). The observed peaks display well-defined Lorentzian lineshapes which correspond to phonons with the energy of few meV and with full-width-at-half-maximum = 0.33 meV \cite{Thouin-19}.
}

Like bipolarons, TI excitons, being bosons, can experience Bose condensation which was predicted in papers  \cite{lit24,lit25}. As distinct from the bipolaron Bose gas, to which a statistically equilibrium description is suitable, for the exciton gas occurring in a quasiequilibrium photoexcited state, this description can be applicable only for long-living exciton states which can be realized in semimetals, gapless semiconductors, systems of nanodots or indirect semiconductors.

In paper  \cite{lit8} exciton condensation was probably observed in a semimetal compound  TiSe$_2$. Since the TI-exciton Hamiltonian  $(\ref{eq4})$ is similar to the bipolaron one, all the results obtained in the statistical description of the TI-bipolaron gas are valid for the case of the TI-exciton gas. In particular, for the temperature of Bose condensation of the TI-exciton gas, we get  \cite{lit22}:
\nonumber
\begin{align}
&T_c(\omega_0)=\Big(F_{3/2}(0)/F_{3/2}(\omega_0/T_c)\Big)^{2/3}T_c(0),
\\
\nonumber
&T_c(0)=3.31\hbar^2n^{2/3}_\text{exc}/M, \quad F_{3/2}(x)=\frac{2}{\sqrt\pi} \int^\infty_0\frac{t^{1/2}dt}{e^{t+x}-1},
\end{align}
where $n_\text{exc}$---is the concentration of TI excitons. Accordingly, a phase transition to the Bose-condensate exciton phase should be of the second order with heat capacity jumps during the transition.

\section{Conclusions}
\label{sec:7}
In this paper, we answered the fundamental question of the role of polaron effects in exciton physics. Though the important role of EPI for excitons in polar media was revealed in a lot of experiments, the question of why the hydrogen-like model appears to be valid under these conditions has remained open thus far \cite{lit26}. In this paper, we have shown that, in the case of exciton-phonon interaction described by Fr\"{o}hlich Hamiltonian, the hydrogen-like model turns out to be applicable if the energy of a transition to the excited state does not exceed the energy of an optical phonon.

The Pekar-Fr\"{o}hlich polaron model is an essential component of a wide range of problems concerned with description of the properties of a particle interacting with a boson reservoir. This model, which was initially introduced to describe the behavior of electrons interacting with phonons in crystals, has been used in such different fields as strongly correlated electron systems, quantum information, high energy physics. Recently, it has been actively used to describe impurity atoms embedded into the Bose-Einstein condensate of ultra cold atoms. The results obtained here, in particular, explain a clearly discernible structure of highly excited (Rydberg) atoms surrounded by the Bose condensate \cite{lit27}.

In conclusion it should be noted that both in the polaron theory and the theory of an exciton interacting with phonons, it is widely believed that self-trapped polaron and exciton states are possible. Thus, for example, by analogy with a polaron, self-trapped exciton states were considered in \cite{lit28,lit29,lit30,lit31,lit32,lit33,lit34,Toyozawa-03}. It was thought there that as the EPI constants exceed a certain critical value, an exciton is trapped by a self-consistent potential produced by it which leads to a possible annihilation of the electron and the hole and disappearance of the exciton. It was also believed that, in the case of a very strong EPI, the energy of the lattice deformed by the exciton can exceed the energy of excitons in a rigid lattice. A change in the energy of such deformed excitons, being negative with respect to excitons in a rigid lattice, can make advantageous spontaneous formation of excitons in crystals with a small gap value, for example, in gapless polar semiconductors (exciton matter  \cite{lit5},\cite{lit35}). {
In this connection notice that there are a lot of works devoted to excitons trapped on defects which interact with optical phonons (i.e., in systems with broken translation symmetry. For review see \cite{Adamowski-94}). It is found that the coupling with phonons always lowers the energy for the delocalized systems, that is, biexcitons and bipolarons. However, for the localized systems, this coupling can lead to the decrease as well as to the increase of the binding depending on material parameters.
}

The results obtained in the paper rule out the possibility of the formation of self-trapped exciton states in translation invariant systems. The conclusions of the possibility of self-trapped excitons in them are based either on an improper choice of probe variation functions, or on erroneous calculations with the use of such functions.

\end{document}